\documentclass[prd,10pt,aps,twocolumn,notitlepage,amsmath,nofootinbib,superscriptaddress,altaffilletter,showpacs,showkeys]{revtex4-2}

\usepackage{tikz} 
\usetikzlibrary{arrows,math}

\usepackage{amsfonts}

\usepackage{graphicx}

\usepackage{hyperref}
\hypersetup{
    colorlinks=true,
    linkcolor=red,     
    citecolor=blue,     
    urlcolor=red       
}
\usepackage[english]{babel}

\begin{document}

\title{ Wind-Finslerian  structure of black holes}

\author{Hengameh R. Dehkordi}\email{hengameh.r@ufabc.edu.br}
\author{Maur\'icio Richartz}\email{mauricio.richartz@ufabc.edu.br}
\affiliation{Centro de Matemática, Computação e Cognição, Universidade Federal do ABC, 09280-560, Santo André, SP, Brazil}

\author{Alberto Saa}\email{asaa@ime.unicamp.br}
\affiliation{Departamento de Matem\'atica Aplicada, Universidade Estadual de Campinas, 13083-859 Campinas, SP, Brazil}

\begin{abstract}
Recently, there has been an increasing interest in the Finslerian interpretation of null geodesics in the exterior regions of stationary black holes, particularly through the Zermelo navigation problem and the Randers metric. In this work, we show that recent mathematical advancements in
  wind-Finslerian structures, which involve 
 the critical and strong Zermelo navigation problems and their connections to Kropina and Lorentz-Finsler metrics, enable the extension of the Finslerian framework to encompass horizons and their interior regions of black holes. The Finslerian indicatrix, a key element of this   framework, serves as an effective tool for identifying frame-dragging effects and the location of horizons and ergosurfaces. We illustrate our results with explicit   physical examples, focusing on 
  spherically symmetric black holes, Kerr black holes, and analog models of gravity. 
Our findings provide new insights into the “river model” of black holes, offering enhanced visual representations of null geodesics on the ergosurfaces, horizons, and within their interior regions.
\end{abstract}
\date{\today}
\maketitle

\section{Introduction}

Finsler geometry is a classical and well-established branch of Mathematics with a vast body of literature. Recently, it has experienced significant developments and has found a wide range of applications in Physics and other applied sciences. While its main foundational ideas were already present in Riemann's {\em Habilitationsvortrag} of 1854, its origins are typically traced back to Paul Finsler's pioneering dissertation \cite{Finsler}. A modern perspective on the topic can be found, for instance, in \cite{Shen}.
For an abbreviated  historical account, see \cite{Chern}.
 Roughly, Finsler geometry is a  Riemannian geometry in which the element of arc length $ds = F(\mathbf{x},d\mathbf{x})$ is not subjected to the usual quadratic restriction $ F^2(\mathbf{x},d\mathbf{x}) = a_{ij} dx^i dx^j$, with $a_{ij}$ standing for the ordinary Riemannian metric
 tensor. 
This restricted case has been historically referred to simply as  Riemannian geometry, and any other form of $F(\mathbf{x},d\mathbf{x})$ is now termed Finslerian.  Numerous Riemannian geometrical notions, ideas, and quantities have corresponding counterparts in Finsler geometry, with virtually every problem in Riemannian geometry having an equivalent formulation in the Finslerian framework.

Arguably, the simplest explicit example of a Finsler geometry in Physics is the Randers spacetime \cite{Randers} with its ``eccentric metric'' now known as the Randers metric
\begin{equation}
\label{Randers}
ds = F(\mathbf{x},d\mathbf{x}) = \sqrt{a_{\mu\nu}(x^\sigma)dx^\mu dx^\nu} + b_\mu(x^\sigma)dx^\mu,
\end{equation}
where  $\mathbf{x}$ is a point with coordinates $(x^\sigma)$  in the spacetime manifold $\mathcal{M}$ and   
 $d\mathbf{x} \in T_\mathbf{x}\mathcal{M}$. We denote the components of $d\mathbf{x}$ by $(dx^\mu)$,
  following the standard physical convention in which Greek indices run over all $n+1$  spacetime coordinates, while Latin indices are limited to the $n$  spatial dimensions. 
 In the Randers metric (\ref{Randers}), $a_{\mu\nu}$ is a standard    metric tensor (Lorentzian, in the case
of Spacetime Physics), and $b_\mu  $ is a one-form (covariant vector) that encodes a directional dependence of the geometry. 
If $b_\mu=0$, the underlying geometry reduces to the standard Riemannian/Lorentzian one associated with the metric tensor $a_{\mu\nu}$. 
Notice that $F(\mathbf{x},d\mathbf{x}) = F(\mathbf{x},-d\mathbf{x})$ for generic vectors
{$d\mathbf{x} \in T_\mathbf{x} \mathcal{M}$}
if, and only if, $b_\mu = 0$.
 In  other words,
a  Randers  metric is symmetric if and only if it is Riemannian. 
Notwithstanding, generic Finsler metrics are positively homogeneous of degree one, {\em  i.e.},
 $F(\mathbf{x},\lambda d\mathbf{x}) = \lambda F(\mathbf{x},d\mathbf{x})$ for all $\lambda \ge  0$.
All Finslerian-geometric objects derived from (\ref{Randers}), such as the fundamental form, connections, covariant derivative, and curvature, are defined starting from $F^2(\mathbf{x},d\mathbf{x})$  analogously to the Riemannian case (see \cite{Shen} and references therein for further details). Positive definite
metrics of the Randers type (\ref{Randers}) arise naturally~\cite{Shen1} in the classical  optimization problem known as the Zermelo navigation problem~\cite{Zermelo}, which consists in finding minimal (non-relativistic) time trajectories in curved spaces under the influence of vector drifts (winds). Many recent advancements in Finsler geometry and its applications in Physics are closely related to the Zermelo problem and its solutions involving  metrics such as those of Randers type. 
  The central mathematical concept explored in the present paper is the notion of a wind-Finslerian structure, which provides a unified model for the classical Zermelo navigation problem across different types of drifts. This concept  was introduced and developed in \cite{miguel}. To be more precise, in all particular cases considered here, the indicatrices are ellipsoids and, consequently, we are dealing with the specific class of wind-Riemannian structures, according to the definitions of \cite{miguel}. In this paper, we will further explore the wind-Finslerian structures and their connections to concrete examples in Spacetime Physics. The reference \cite{miguel} also provides a comprehensive review of the recent mathematical literature on this topic.

  From a physical standpoint, one does not expect distinguished local spacelike directions.  However, this does not apply to timelike directions; in many circumstances, the future and the past causal cones are not merely related by simple inversions.  In fact, this was the main motivation behind Randers' introduction of a  spacetime   metric 
with asymmetrical properties 
of the type (\ref{Randers}) in the early forties \cite{Randers}. Moreover, the Principle of Equivalence imposes restrictions on possible forms of a Finsler metric from  the  Spacetime Physics perspective \cite{Bekenstein}. 
Interestingly, positive definite metrics of type (\ref{Randers})
often appear in specific spacetime problems restricted to spatial sections,  such as the description of causal structures.  In particular, the propagation of light rays in stationary spacetimes can be effectively described by a Randers spatial metric. These issues were mathematically 
established in \cite{Stationary} and then considered from a physical point of view for black holes in \cite{gibbons2009stationary}. 
The extension for nonstationary but 
 stably causal spacetimes admitting a globally defined time coordinate  was considered in \cite{App0}.
 More recently, \cite{LiJia} has shown that the Zermelo problem can be used to generate spacetimes with specific causal properties. To date, only the propagation of light rays in the exterior region of black holes 
  with a non-vanishing Killing vector 
  has been described in Finslerian terms
 in the  Physics  literature. Extending the Finsler structure across the event horizon 
is a physically relevant problem, as recognized in \cite{LiJia}.  
    Notably, recent research in the Mathematics literature has tackled these issues. For instance, \cite{Conformal} shows that there exists a one-to-one correspondence between Randers spaces and conformal classes of stationary spacetimes, with the causal structure of the latter fully encoded in the metric properties of the former. Furthermore, the primary objective of Ref.~\cite{miguel} was to extend
   the
Finslerian description of causality and horizons in any region of a spacetime admitting a spacelike slice and a non-vanishing transversal Killing
vector field.  One of the main goals of the current study is precisely to establish a connection between these recent mathematical findings and the physical literature. 
We highlight also that
recent applications of Finsler geometry in other problems of General and Special Relativity
can be found, {\em  e.g.,} in \cite{App1,App2,App3,App4,App5,App6,App7,App8,App9}.

In this paper, we demonstrate how recent mathematical progress~\cite{miguel,Stationary,Conformal,Krop1,Krop2,LF} can be explored to interpret  light propagation within the interior region of black holes in Finslerian terms. The key ingredients are the Kropina
and Lorentz-Finsler metrics, which appear in certain extended versions of the Zermelo navigation problem. We show, through explicit  and  physically relevant examples, that the Kropina metric adequately describes the causal properties of ergosurfaces in stationary black holes, while the Lorentz-Finsler metric captures, in the region where it is well defined, the causal structure of the interior (non-stationary) region of black holes. In this way, we show that it is indeed possible to have, based on recent mathematical developments, a complete Finslerian description of the causal structure of black holes, shedding new light on phenomena such as the one-directional membrane behavior of horizons~\cite{Membrane} and the river model paradigm of black holes~\cite{river}. 
 
 The paper is organized as follows. In the next section, we review recent mathematical developments related to the Kropina and Lorentz-Finsler metrics and compile the results that are instrumental in describing the causal structure of black holes, which is explored in Section \ref{sec3}.   In Sec.~IV, we analyze analog models of gravity within the Finslerian framework.  The final section is dedicated to 
our concluding remarks, which include brief comments on Fermat’s principle  and analog spacetimes. 
 We adopt $G=c=1$ units throughout this work.

\section{Finsler  geometry and the Zermelo problem}

We begin by revisiting the classical Zermelo navigation problem in curved space, as it provides the essential foundation for our approach in Section \ref{sec3}. Consider a non-relativistic particle moving with constant (unit) speed in an $n$-dimensional Riemannian manifold $(\mathcal{M}, h_{ij} )$ subject to a time-independent vector field drift (wind) $W^i(x^j)$ such that $||W||^2 = h_{ij}W^iW^j < 1$ (weak drift). The Zermelo problem consists in finding the trajectory that minimizes the transit time for a particle traveling between two distinct points of $\mathcal{M}$. This minimal time trajectory corresponds to the geodesic of a positive definite Randers metric (\ref{Randers}) with~\cite{Shen1}
\begin{equation}
\label{RandCond}
a_{ij} = \frac{1}{\lambda^2}\left(\lambda h_{ij} + W_iW_j \right)  \quad {\rm and}\quad 
b_i = -\frac{W_i}{\lambda},
\end{equation}
where $W_i = h_{ij}W^j$ and
$
\lambda = 1 - ||W||^2
$. The weak drift condition, which implies $\lambda > 0$, is essential to this derivation.
It is worth noticing~\cite{gibbons2009stationary} that 
\begin{equation}
\label{b2}
||b||^2_F = a^{ij}b_ib_j =  ||W||^2,
\end{equation}
where $a^{ij}$ is the inverse of the metric tensor $a_{ij}$. Hence, the weak drift also implies $||b||^2_F < 1$, which is commonly known 
  as the Finsler condition in the Physics literature. The so-called Randers $(a_{ij},b_j)$ and Zermelo $(h_{ij},W^j)$ data are connected through a Legendre transformation \cite{gibbons2009stationary}.
As it will become clear in the next section, describing the  wind-Finslerian  structure of black holes requires extending the Zermelo problem to the critical $(||W||=1)$ and strong $(||W||>1)$ drift cases. Fortunately, both problems have been thoroughly investigated and, for the purposes of this work, conclusively solved in the recent mathematical literature.   Ref.~\cite{miguel}, which first solved the strong drift case, also provides a useful review of the pertinent mathematical literature.  

 The Zermelo problem for a critical drift was solved in \cite{Krop1,Krop2} through the analysis of geodesics in the singular limit $\lambda = 1$ of the Randers metric (\ref{Randers}). The resulting metric
\begin{equation}
\label{Kropina}
 F(\mathbf{x},d\mathbf{x}) =  \frac{h_{ij}dx^idx^j}{2W_jdx^j},
\end{equation}
   defined
on the half-tangent space 
\begin{equation}
\label{halfspace}
A = \{d\mathbf{x}  \in T_\mathbf{x}\mathcal{M}\ :\  W_jdx^j > 0\},
\end{equation}
is referred to as the Kropina metric in the mathematical literature. 
 Here, the germ of the one-directional membrane behavior which characterizes horizons  becomes apparent: solutions (minimal time trajectories) exist only when their tangent vectors $d\mathbf{x} = (dx^i)$ are positively aligned with the drift $W^j$. 
 
 The strong drift case, on the other hand, was solved in \cite{miguel},   where it was shown that minimal time trajectories 
are associated with  
 the geodesics of the minus-sign case  of the so-called Lorentz-Finsler metrics
\begin{equation} 
\label{LF}
 F(\mathbf{x},d\mathbf{x}) =- \frac{1}{\lambda}\left(W_idx^i \mp\sqrt{\left(W_jdx^j\right)^2 +\lambda h_{ij}dx^idx^j}  \right), 
\end{equation}
defined on the conic region of the tangent space $T_\mathbf{x}\mathcal{M}$ given by
\begin{eqnarray}
\label{cone}
 A=& &\left\{  \mathbf{x}\in T_\mathbf{x}\mathcal{M}:  \phantom{\left(W_jdx^j\right)^2>0}\right.
 \nonumber \\ 
 && \left. \lambda h_{ij}dx^idx^j + \left(W_jdx^j\right)^2>0 ,\  W_jdx^j > 0 \right\}.
\end{eqnarray}
We shall refer to the minus and plus cases, respectively, as first and second Lorentz-Finsler metrics. The strong drift
condition  $\lambda < 0$  
 guarantees that both metrics in (\ref{LF}) are positive   functions on $A$. Additionally, as in the critical case, the problem exhibits a one-directional behavior: a solution exists only if the trajectories are positively aligned with the drift, and moreover their instantaneous velocities lie within the conic region (\ref{cone}).  It is worth noticing that the transition from weak to strong drift involves several mathematical subtleties. We will adopt a more heuristic approach in the next subsection. The reader can find the pertinent mathematical details in Ref. \cite{miguel}.

The unit ball on $T_\mathbf{x}\mathcal{M}$ defined from the condition $F(\mathbf{x},d\mathbf{x} )=1$, called the indicatrix of the Finsler metric, encodes important information on the geodesics. In particular, the indicatrix determines the possible directions 
$d\mathbf{x}\in T_\mathbf{x}\mathcal{M}$ 
 of geodesics starting at the point $\mathbf{x}\in\mathcal{M}$. In our case,
for all metrics, including the second Lorentz-Finsler, the
indicatrices are given by
\begin{equation}
\label{ind}
h_{ij}(dx^i-W^i)(dx^j-W^j) = 1.
\end{equation}
In other words, the indicatrices of the Finsler metrics associated with the Zermelo navigation problem on a Riemannian manifold
$(\mathcal{M},h_{ij})$ with drift $W^i$, regardless of the value of $||W||$, correspond to the unit balls of the
Riemannian metric $h_{ij}$, with their centers shifted by the vector $W^i$. Randers had already recognized this type of eccentricity in his pioneering work \cite{Randers}.  
The use of displaced indicatrices as tools
to describe causal and geometric structure, an important result of Ref. \cite{miguel}, 
 will be essential for the analysis of
the causal structure of black holes in Section \ref{sec3}.

\subsection{Geodesic matching}

The mathematical description of the causal structure of Finsler spaces is both comprehensive and sophisticated, see, {\em e.g.},~\cite{miguel,Causal1,Causal2,Ettore1, Ettore2}. For our purposes here, we
need only to consider the matching of geodesics across the three cases of Zermelo's problems,  
whose general validity is assured by the rigorous mathematical results 
on  wind-Finslerian structures introduced in \cite{miguel}. 
 Differentiability issues will not be discussed, as all functions are assumed to be sufficiently smooth. 
 Let $W^i$ be a generic smooth vector field drift on a Riemannian manifold  $(\mathcal{M}, h_{ij} )$  and  assume there are regions of $\mathcal{M}$ characterized by weak ($||W||<1$) and strong $(||W||>1)$ drifts. The   boundary between these regions
  corresponds to the critical drift case $||W||=1$, and in the physical cases where it is associated with a
horizon, it will be a co-dimension 1 
 submanifold $\mathcal{H}$ of $\mathcal{M}$.  
  The key observation is that geodesics can cross smoothly from the weak to the strong drift regions,  provided that the half-space (\ref{halfspace}) and
 conic (\ref{cone}) constraints are satisfied in the pertinent regions.
  This is possible because the Randers (\ref{Randers}), Kropina (\ref{Kropina}), and first Lorentz-Finsler (\ref{LF}) metrics all give rise to the same Finslerian Lagrangian action   $S=\int F(\mathbf{x},d\mathbf{x} )$ for geodesics,  and they can be smoothly matched across the drift region transitions.    In fact, the three cases can be nicely unified if one uses the Finslerian Lagrangian action  with
\begin{equation}
 F(\mathbf{x},d\mathbf{x}) =  \frac{h_{ij}dx^idx^j}{ W_idx^i +\sqrt{\left(W_jdx^j\right)^2 +\lambda h_{ij}dx^idx^j} } , 
\end{equation}
which is valid for any value of $\lambda$, 
see \cite{miguel} for further details. 
 Notice that the   half-space and
 conic  constraints are essential to assure that we have   positive Lagrangians $F(\mathbf{x},d\mathbf{x} )$ and consequently  well-defined geodesic problems, on $\mathcal{H}$ and in the strong drift region. 
 
A concrete example is instructive here. Although simple, the following scenario is paradigmatic, since it corresponds to the river model of the Schwarzschild black hole~\cite{river}, which we discuss in the next section. Let us examine the three-dimensional Euclidean space $(\mathbb{R}^3,\delta_{ij})$
with standard Cartesian coordinates $(x,y,z)$, endowed with an infalling radial drift such that $W_idx^i=-\sqrt{2M/r} \, dr$, where $r^2 = x^2 + y^2 + z^2$.
In this case, $\mathcal{H}$ is a spherical surface of radius $r=2M$, identified with the event horizon of the black hole. Its interior and exterior correspond, respectively, to the strong and weak drift regions.

Given the spherical symmetry of the problem, we can consider, without loss of generality, the 
   Randers (\ref{Randers}), Kropina (\ref{Kropina}), and the first Finsler-Lorentz (\ref{LF}) metrics along the $x$ axis.
A unified description is given by 
\begin{equation}
\label{metr1}
F(\mathbf{x},d\mathbf{x}) = \frac{1}{\lambda}\left(\sqrt{dx^2 + \lambda ( dy^2+dz^2)} + \sqrt{\frac{2M}{r}} dx \right),
\end{equation}
with 
$\lambda = 1 - 2M/r$.
The three metrics   correspond, respectively, to 
 $r>2M$ ($\lambda >0$), $r=2M$ ($\lambda = 0$), and $r<2M$ ($\lambda <0$). Hence,   a unique Finslerian Lagrangian is defined over the entire range of $r$, which implies that geodesics can smoothly cross the submanifold $r=2M$ provided the half-space (\ref{halfspace})    and conic (\ref{cone}) constraints are satisfied in the respective regions.  
 
\begin{figure}[t]
\begin{tikzpicture}[
    thick,  scale=1.2, 
    >=stealth',
    dot/.style = {
      draw,
      fill = white,
      circle,
      inner sep = 0pt,
      minimum size = 4pt
    }
  ]
  
 \draw[->] (0,0) -- (7,0) coordinate[label = {below:$x$}]  ;

\draw[red,densely dashed] (4,-1.6) -- (4,1.5) ; 

\node  at (4.,1.9) {\footnotesize $r=2M$};
\node  at (4.,1.6) {\footnotesize (event horizon)};
    
\draw (3.5,0) circle (.5) ; 

\draw[->,red] (4,0) -- (3,0);

\draw[->,red] (4,0) -- (3.66,0.47);
\draw[->,red] (4,0) -- (3.66,-0.47);
\draw[->,red] (4,0) -- (3.33,0.47);
\draw[->,red] (4,0) -- (3.33,-0.47);
\draw[->,red] (4,0) -- (3.08,0.28);
\draw[->,red] (4,0) -- (3.08,-0.28);

\filldraw  (4,0) circle (0.05);


\draw (6,0) circle (.5) ; 

\draw[->,red] (6.15,0) -- (6.5,0);
\draw[->,red] (6.15,0) -- (5.5,0);

\draw[->,red] (6.15,0) -- (5.67,0.37);
\draw[->,red] (6.15,0) -- (5.67,-0.37);

\draw[->,red] (6.15,0) -- (6,0.5);
\draw[->,red] (6.15,0) -- (6.0,-0.5);

\draw[->,red] (6.15,0) -- (6.38,0.33);
\draw[->,red] (6.15,0) -- (6.38,-0.33);

\filldraw  (6.15,0) circle (0.05);


\draw (1,0) circle (.5) ; 

\draw[->,red] (2,0) -- (0.5,0);
\draw[->,red] (2,0) -- (1.5,0);

\draw[->,red] (2,0) -- (0.83,0.47);
\draw[->,red] (2,0) -- (1.45,0.22);
\draw[->,red] (2,0) -- (0.83,-0.47);
\draw[->,red] (2,0) -- (1.45,-0.22);

\draw[->,red] (2,0) -- (0.59,0.28);
\draw[->,red] (2,0) -- (0.59,-0.28);

\draw[->,red] (2,0) -- (1.25,0.43);
\draw[->,red] (2,0) -- (1.25,-0.43);

\filldraw  (2,0) circle (0.05);

\draw[dotted,blue] (2,0) -- (0.5,0.86);
\draw[dotted,blue] (2,0) -- (0.5,-0.86);

\end{tikzpicture}
    \caption{Depiction of the Finslerian indicatrices for the three drift regimes of the Schwarzschild case along the $x$ axis in the equatorial plane. In all cases, the indicatrices are unit spheres, with their centers displaced in the $x$ direction according to  (\ref{indica}). Right: The weak drift regime $r>2M$, in which null geodesics starting from a given point can propagate in all directions. Center: The critical regime $r=2M$, in which the propagation of null geodesics is restricted to the directions that are positively aligned with the drift according to the half-space restriction (\ref{halfspace}). In this case, the half-space is tangent to the critical submanifold
$\mathcal{H}$    
     (the event horizon),
implying that there is no outgoing null geodesics starting on $\mathcal{H}$, establishing  its causal disconnection from the exterior region.
 Left: The strong drift regime. Here, propagation is similarly limited to directions positively aligned with the drift and, moreover, constrained to lie within the conic region (blue dotted lines) according to (\ref{cone}). Notice that as the starting point of the geodesics approaches the singularity at $r=0$, the displacement of the center of the unit sphere increases, resulting in more acute cones. In the vicinity of the central singularity, all geodesics are rapidly dragged towards it, and there is no way to  prevent them from reaching $r=0$,   in perfect agreement with the understanding that $r=0$ is a future timelike singularity.  This is a concrete realization, for 
a  Schwarzschild black hole, 
  of the wind-Finslerian description 
 of the causal structure  of spacetimes introduced in \cite{miguel}.  
    }\label{fig1}
\end{figure}
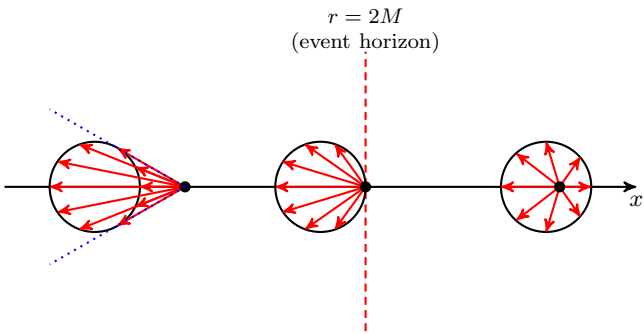

 The Finslerian indicatrix (\ref{ind}) associated with (\ref{metr1}) reduces to the simple expression
\begin{equation}
\label{indica}
\left( dx + \sqrt{\frac{2M}{r}}\right)^2 + dy^2 + dz^2=1,
\end{equation}
which can be readily identified as the unit sphere with a displaced center. Notice that (\ref{indica}) is valid
for all metrics, including the second Lorentz-Finsler, and we have
\begin{equation}
\label{dx}
-1 - \sqrt{\frac{2M}{r}} \le dx \le 1 - \sqrt{\frac{2M}{r}},
\end{equation}
for all cases. 
In the exterior (weak drift) region, characterized by $ 
\sqrt{2M/r}<1$, both ingoing ($dx <0$) and outgoing ($dx > 0$) geodesics are allowed.
For  $r\to \infty$, the indicatrix sphere is centered at the origin, implying that there is no preferred direction, as one would expect in asymptotically flat cases. 
 On the other hand, on the critical submanifold  $r=2M$,  there are no outgoing geodesics, as $-2 \le dx \le 0$. Finally, for the interior (strong drift) region, where $\sqrt{2M/r}>1$, all geodesics are ingoing since $dx<0$. Fig.~\ref{fig1} illustrates the three cases. 
The resemblance of this figure to the river model of black holes~\cite{river} is evident. We emphasize that all information regarding geodesics, including their directional dependence, is encoded in the Finsler metric (\ref{metr1}), which holds for all $r>0$.

\section{Black holes}
\label{sec3}

We now proceed to examine the causal structure of black holes through the Finslerian framework. 
Our first remark is that the forward-in-time ($dt>0$) null geodesics ($ds=0$) of any stationary spacetime whose line element is cast in the Painlev\'e-Gullstrand form~\cite{PG}
\begin{eqnarray}
\label{PG}
ds^2 &=&   g_{\mu\nu}dx^\mu dx^\nu \\
&=& -dt^2 +h_{ij}\left(dx^i - W^idt \right) \left(dx^j - W^jdt \right), \nonumber 
\end{eqnarray}
with  both time-independent spatial metric $h_{ij}$ and drift $W^i$, are determined by a 
positive definite Finsler metric $dt = F(\mathbf{x},d\mathbf{x} )$ of the Randers type characterized by (\ref{Randers}) and (\ref{RandCond}).  
In the standard river model of black holes~\cite{river}, $h_{ij}$ is assumed to be an Euclidean metric, and in this case the metric
(\ref{PG}) is said to be of strong Painlev\'e–Gullstrand form~\cite{visserPG}. We note, however, that this restriction on $h_{ij}$ is not required in our analysis. The metric (\ref{PG}), up to a conformal term, is   referred to as the Zermelo form of a stationary spacetime in the recent   Physics literature~\cite{gibbons2009stationary,LiJia} . 
 We also remark that the spacetimes for which the main recent mathematical results  were established \cite{miguel}, possessing a space-transverse Killing vector field, have metrics precisely of the form (\ref{PG}), with $g_{00}=1-||W||^2$ (note that $g_{00}$ in our notation corresponds to $\Lambda$ in the notation of \cite{miguel}).  
Building on the discussion of Zermelo's problem in Section II, we can describe the  propagation of null geodesics of (\ref{PG}) using the underlying
Finsler metric $dt = F(\mathbf{x},d\mathbf{x} )$, irrespective of the value of $\lambda = 1 - h_{ij}W^iW^j$.  
The value of $\lambda$, in turn, determines whether the Finsler metric takes the form of Randers (\ref{Randers}), Kropina (\ref{Kropina}), or Lorentz-Finsler (\ref{LF}). Accordingly, it also identifies the relevant Zermelo problem associated with the black hole spacetime under investigation. 
We will now consider static and stationary
black holes separately. 

\subsection{Static black holes}

The generic line element for a static spherically symmetrical spacetime reads
\begin{equation}
\label{spherical} 
ds^2 = g_{\mu\nu}dx^\mu dx^\nu = - f(r)d\tau^2 +    \frac{dr^2}{f(r)}+ R^2(r) dS^2,
\end{equation}
where $dS^2$ stands for the usual   metric on the unit sphere. 
Asymptotic flatness requires $R \sim r$ and $f \sim 1$ for sufficiently large $r$. 
The zeros of  $f(r)$ correspond to Killing horizons of this spacetime,
 with the outermost zero corresponding to the event horizon of the black hole. We can cast (\ref{spherical}) in a  Painlevé-Gullstrand form by introducing the new time variable \cite{PG}
\begin{equation}
\label{PG0}
dt  = d\tau + \frac{1}{f}\sqrt{1- f}dr,
\end{equation}
leading to a metric of the form (\ref{PG}) with drift given by
\begin{equation}
\label{drift}
W_idx^i =  -\sqrt{1-f}dr
\end{equation}
and spatial Riemannian metric
\begin{equation}
d\ell^2 = h_{ij}dx^i dx^j = dr^2 +R^2(r)dS^2.
\end{equation}
Note that, unless $R(r)=r$, the associated Zermelo navigation problem is defined on a curved manifold. Additionally, we have $\lambda = f$  
and the underlying Finsler metric reads
\begin{equation}
\label{nullgeo}
dt \! = \! F(x,dx ) \! = \! \frac{1}{f}\left(  \! \sqrt{1-f}  dr  + \sqrt{ dr^2 + fR^2dS^2 } \right) \! . \! 
\end{equation}

The importance of the Painlevé-Gullstrand coordinates becomes evident now. They can be smoothly extended across the horizon ($f=0$), yielding the Finsler metric (\ref{nullgeo}), which remains well defined throughout the entire domain of the radial coordinate $r$ where the Painlevé-Gullstrand 
metric is also defined. Consequently, the Finsler metric (\ref{nullgeo}) is also well defined for all values of $\lambda$.
Besides, in the strong drift case $(\lambda < 0)$, we also have the second Lorentz-Finsler metric.
 We now turn to some explicit examples for illustration.
 
 The Schwarzschild case analyzed in the previous section can be generalized to other black holes described by (\ref{spherical}). In fact, any black hole with metric such that $R(r)=r$ will correspond to a Zermelo problem defined on a flat tridimensional manifold. 
The corresponding indicatrices, determined by (\ref{ind}),  are
also given by spheres with displaced center, such that
\begin{equation}
\label{ellipsoid}
\left(dx + \sqrt{1-f} \right)^2 + dy^2 + dz^2 = 1,
\end{equation}
which implies 
\begin{equation}
\label{dx1}
-1 - \sqrt{1-f} \le dx \le 1 - \sqrt{1-f}.
\end{equation}
 As in the Schwarzschild case, limiting the analysis to the $x$ axis does not result in any loss of generality, owing to the spherical symmetry of the problem.

Many other physically relevant spacetimes belong to this class of Euclidean Zermelo problems. For instance, the Rindler metric 
considered in  \cite{LiJia},  characterized by the constant radial acceleration  $\gamma$, corresponds to
$f=1-2\gamma r$. For the 
  Reissner-Nordstr\"om black hole,  characterized by the electric charge $Q$ such that $|Q|<M$, we have
\begin{equation}
\label{RN}
f(r) = 1-\frac{2M}{r} +\frac{Q^2}{r^2} = \frac{(r-r_-)(r-r_+)}{r^2},
\end{equation}
where 
$r_{\pm} =  M\pm \sqrt{M^2 - Q^2}  $. The Reissner-Nordstr\"om-de Sitter case can also be accommodated here, see \cite{LiJia} for additional examples and further details. We emphasize that there are static black holes, such as dilaton black holes~\cite{dilaton}, which necessarily require a curved spatial manifold. The only difference with respect to the flat cases is that in these non-Euclidean scenarios, the indicatrix  (\ref{ind}) will be an eccentric ellipsoid instead of an eccentric sphere as in (\ref{ellipsoid}). Still, ultimately, the qualitative results will remain the same. 

It is worth examining the Reissner-Nordstr\"om black hole in greater detail.
It is clear from (\ref{RN}) that we have two Killing horizons, 
but there is also  another distinguished radius, namely
$r=r_0= Q^2/2M$. The Painlevé-Gullstrand
metric for the Reissner-Nordstr\"om black hole is defined only for $r\ge r_0$, since $f(r)>1$ for $r<r_0$ and the square root in (\ref{PG0}) fails to be real. 

Given that $r_0$ is half of the harmonic mean of the two horizons, it follows that 
 $r_0 <r_-$. Hence, the Painlevé-Gullstrand coordinates for the Reissner-Nordstr\"om black hole effectively extend across its inner horizon $r=r_-$.
For $r>r_+$,   $\lambda = f(r)>0$ and we are in the weak drift regime of the Zermelo problem. Similarly to the
exterior region of the Schwarzschild black hole of the previous section,  in this region we can have  both  ingoing ($dx<0$) and outgoing ($dx>0$) null geodesics.

For $r=r_+$ and $r=r_-$, we have $f(r)=0$, which correspond to the critical Zermelo case. The associated geodesics are such that $-2\le dx \le 0$, as in the case of the Schwarzschild event horizon.  In the region between the two Killing horizons, $r_- < r <r_+$,  where  $f(r)<0$, we have the strong drift case.
 This regime is similar to the one occurring in the
interior region of the Schwarzschild black hole, where we have only ingoing geodesics since $dx<0$. 
The novelty for the Reissner-Nordstr\"om black hole is that, for $r_0\le r < r_-$, we have again the weak drift regime
 corresponding to 
$f(r)>0$.
 In this regime, both ingoing ($dx<0$) and outgoing ($dx>0$) null geodesics are allowed, but they remain entirely confined within the inner horizon.  
This indicates that null geodesics starting within the inner (Cauchy) horizon of the Reissner-Nordström black hole can escape from reaching the spacelike singularity at $r = 0$. This characteristic is well-known in the theoretical study of charged black holes. However, it is generally not considered physically relevant because the electric charge in astrophysical black holes is expected to neutralize quickly~\cite{Gibbons:1975kk,Neutral}. Additionally, there are inherent instabilities associated with the Cauchy horizon, see \cite{CauchyInstHiscock,CauchyInstPI,CauchyInst}.

We remark that the Painlevé-Gullstrand coordinates for the Reissner-Nordstr\"om black hole are defined only for $ r > r_0$. Consequently, the Finslerian description of its causal structure is limited by the same restriction. However, this does not pose a major concern for our purposes, as it is related to the presence of a negative local mass, which effectively prevents ingoing geodesics from reaching the central singularity at $r = 0$,
a well-known issue with the Painlevé-Gullstrand coordinates. For further details, see \cite{Faraoni}. For an interpretation of this phenomenon in the context of the river model of black holes, refer to \cite{river}.
  Notwithstanding, Finsler geometry offers a fresh and physically insightful view of this issue. The local mass function relevant to the problem is the so-called Misner-Sharp-Hernandez mass $M_{\rm  MSH}(r)$, which plays a central role in spherical fluid mechanics and  gravitational collapse~\cite{MS,HM}. For a  static spherically symmetrical spacetime, this mass is introduced by recasting the function $f$ in (\ref{spherical}) as
\begin{equation}
f(r) = 1 - \frac{2M_{\rm  MSH}(r)}{r},
\end{equation}
this mimicking some properties of the Schwarzschild mass. For a Reissner-Nordstr\"om  black hole, this local mass reads
\begin{equation}
M_{\rm  MSH}(r) = M - \frac{Q^2}{2r},
\end{equation}
implying that $M_{\rm  MSH}$ is non-positive for $0<r\le r_0$, thus resulting in a non-attractive behavior that prevents all geodesics from reaching the center $r=0$. Remarkably, when expressed in terms of the Misner–Sharp–Hernandez mass, the Finsler metric (\ref{nullgeo}) for generic static, spherically symmetric spacetimes with $R(r) = r$, along with the associated indicatrix expressions (\ref{ellipsoid}) and (\ref{dx1}), reduce to those of the Schwarzschild case (\ref{metr1}), (\ref{indica}), and (\ref{dx}), with the Schwarzschild mass replaced by $M_{\rm  MSH}(r)$.  This shows that $M_{\rm  MSH}(r)$ does more than mimic the Schwarzschild mass
in the underlying Finsler geometry of a static spherically symmetric black hole; it fully encodes all the Finsler geometrical
properties, offering a new geometric perspective in the Misner-Sharp-Hernandez mass. 

Yet, regarding the issue of geodesic repulsion in the region $0<r\le r_0$, one should recall the Martel-Poisson family of coordinates \cite{PG, Faraoni}, which can effectively describe free-fall motion for $r<r_0$ and could thus be explored to extend the  wind-Finslerian  structure of black holes to regions
with negative Misner-Sharp-Hernandez mass. Such coordinates correspond to the choice
\begin{equation}
\label{PG1}
dt  = d\tau + \frac{1}{f}\sqrt{1- pf}dr,
\end{equation}
instead of (\ref{PG0}), with constant $p$ such that $0 < p \le 1$. 
For the physical interpretation of the parameter $p$, see \cite{PG, Faraoni}. Notice that $p=1$ corresponds to the usual 
Painlev\'e-Gullstrand coordinates  and the (singular) limit $p\to 0$ is related to the  Eddington-Finkelstein coordinates \cite{Faraoni}. 
Inserting (\ref{PG1}) in (\ref{spherical}) 
and looking for   forward-in-time   null geodesics, one obtains the Finsler metric
\begin{equation}
\label{nullgeo1}
\! dt \! = \! F(x,dx ) \! = \! \frac{1}{f}\left(  \! \sqrt{1-pf}  dr  + \sqrt{ dr^2 + fR^2dS^2 } \right) \!  ,\!  
\end{equation}
which,  in the Reissner-Nordstr\"om case, is associated to the ellipsoidal indicatrix
\begin{equation}
\label{ellipsoid2}
p\left(dx + \frac{\sqrt{1-pf}}{p} \right)^2 + dy^2 + dz^2 = \frac{1}{p},
\end{equation}
implying
\begin{equation}
\label{dx2}
-1 -\sqrt{1-pf} \le p \, dx \le 1 -\sqrt{1-pf}.
\end{equation}
From (\ref{b2}) and (\ref{nullgeo1}), we see that the critical drift region is unaltered by $p$; it always corresponds
to the zeros of $f(r)$, and hence the
qualitative behavior of the null geodesics obeying (\ref{ellipsoid2}) and (\ref{dx2}) remains the same regardless of the
value of the constant $p$.  
 However, the Finsler structure associated with the Martel-Poisson family of coordinates is well defined for any 
$r> r_0^*$, with $r_0^*$ being the solution of
\begin{equation}
\frac{2M_{\rm  MSH}(r)}{r} = - \frac{1-p}{p}.
\end{equation}
Note that $r_0^*$ lies within the region $0< r\le r_0$ of geodesic repulsion, where $M_{\rm  MSH}$ is negative. Finsler geometry, consequently, plays a significant role in understanding the internal region of static black holes, enlightening the relationship between the Misner-Sharp-Hernandez local mass function and the causal structure of the spacetime.

\subsection{Stationary black holes}

The case of stationary black holes is significantly richer and physically more interesting. Here, we focus exclusively on the Kerr metric, which describes electrically neutral rotating black holes, thus avoiding unnecessary complications such as the negative local mass of Reissner-Nordstr\"om black holes discussed above.

It is well known that the Kerr metric cannot be cast in 
a strong Painlev\'e–Gullstrand form~\cite{visserPG}. In fact, its representation within the river model paradigm relies on the Doran coordinates~\cite{Doran}, requiring the introduction of a ``twist'' drift. In this work, we consider the Painlev\'e–Gullstrand form of the Kerr metric proposed by Natario in \cite{Natario}, which corresponds to the Zermelo metric (\ref{PG}) with 
the Riemannian spatial metric
\begin{equation}
\label{Nat}
h_{ij}dx^idx^j = \frac{\rho^2}{\Sigma}dr^2 + \rho^2d\theta^2 + \Sigma\sin^2\theta \left(d\phi + \delta d\theta \right)^2,
\end{equation}
and drift $W^i$ such that
\begin{equation}
\label{DriftNat}
W_idx^i= \frac{\rho^2}{\Sigma} v dr + \Sigma\sin^2\theta \Omega d\phi,
\end{equation}
with 
\begin{equation}
v = -\frac{\sqrt{2Mr(r^2+a^2)}}{\rho^2}, \quad \Omega = \frac{2Mra}{\rho^2\Sigma},
\end{equation}
\begin{equation}
\rho^2 = r^2 + a^2\cos^2 \theta,\quad
\Sigma = r^2 + a^2 + \frac{2Mra^2}{\rho^2}\sin^2\theta,
\end{equation}
and
\begin{equation}
\delta = -a^2  \sin( 2\theta) \int_r^{\infty} \frac{v\Omega}{\Sigma}dr.
\end{equation}
We remark that the Kerr metric describes a black hole, spinning with angular velocity $\Omega$, only if $|a|\le M$; otherwise it corresponds to a naked singularity.

Notice that (\ref{Nat}) and (\ref{DriftNat}) are well defined for all positive $r$. 
Natario's coordinates $(t,r,\theta,\phi)$ differ from the usual Boyer-Lindquist in the time   variable, which is not relevant to our analysis, 
and in the azimuthal angle. The usual  Boyer-Lindquist azimuthal coordinate $\phi_{BL}$ is given by (see \cite{Natario} for details)
\begin{equation} \label{phiBL}
d\phi_{BL} = d\phi +\frac{\rho^2}{\Delta}v\Omega dr+  \delta d\theta,
\end{equation}
where $\Delta = r^2 -2Mr + a^2$. 
Since the coordinates $r$ and $\theta$ are the same in Boyer-Lindquist and Natario's coordinates, together with the fact (\ref{Nat}) and (\ref{DriftNat}) do not depend on $t$ and $\phi$ (as expected due to axisymmetry and stationarity),
the ergosurfaces and the horizons of the Kerr metric are described by the same equations in both coordinate systems. In fact, the outer and inner ergosurfaces are given by the solutions of $\rho^2 = 2Mr$, whereas the outer and inner event horizons are given by the roots of $\Delta = 0$.

 The Finsler indicatrix for (\ref{Nat}) and (\ref{DriftNat}) reads
\begin{equation}
\label{KerrInd}
\frac{\rho^2}{\Sigma}\left(dr-v \right)^2 +\rho^2d\theta^2 + \Sigma\sin^2\theta(d\phi +\delta d\theta - \Omega)^2 = 1.
\end{equation} 
Geometrically, it
corresponds  to an ellipsoid with displaced center in the tangent space of the spatial sections of (\ref{PG}) such that
\begin{equation}
-\frac{\sqrt{\Sigma}}{\rho}\left( 1 - \frac{\rho v}{\sqrt{\Sigma}} \right)\le dr  \le \frac{\sqrt{\Sigma}}{\rho} \left( 1 + \frac{\rho v}{\sqrt{\Sigma}}\right).
\end{equation}
Given the relation
\begin{equation}
\frac{\rho v}{\sqrt{\Sigma}} = -\sqrt{1 -\frac{\Delta}{\Sigma}},
\end{equation}
we deduce that $dr\le 0$ when $\Delta=0$, which implies  that no geodesic can be outgoing at either horizon. Between the horizons, the 
equations above imply
$dr < 0$, meaning all geodesics are ingoing. Hence, the qualitative behavior of the event horizons of a Kerr black hole is analogous to what was observed previously for static black holes. 
Additionally, similarly to the Reissner-Nordstr\"om case,
null geodesics can avoid the ring singularity at $r = 0$  when $0\le r \le r_-$, although they remain confined to the interior of the inner horizon, with the same caveats concerning the physical relevance of this picture due to   instabilities issues of the inner horizon \cite{CauchyInst}.

The analysis of the ergosurfaces is more involved. Consider a geodesic which is locally aligned with the drift vector $W^i$, characterized by
\begin{equation}
\label{align}
dr = \alpha v, \quad d\theta =0, \quad d\phi = \alpha \Omega.
\end{equation}
In this case, the equation (\ref{KerrInd}) that determines the indicatrix simplifies to
\begin{equation}
\label{eq}
(\alpha -1)^2 ||W||^2 = 1,
\end{equation}
where
\begin{equation}
||W||^2  = \frac{\rho^2 v^2}{\Sigma}+ \Sigma\sin^2 \theta \Omega^2 = \frac{2Mr}{\rho^2}.
\end{equation}
From this relation, we see that the 
 ergosurfaces correspond effectively to the critical drift case $||W||=1$.  Since the solutions of (\ref{eq}) are $\alpha = 1 \pm ||W||^{-1}$, we have that on the ergosurfaces we effectively always have a  blocked direction, 
{\em i.e.}, a direction such that  $dx^i=0$ for null geodesics.   More precisely, the  blocked  propagation direction   for null geodesics  
corresponds to the origin of $T_\mathbf{x}\mathcal{M}$, and the origin belongs to the indicatrix only   in the critical case. 
However, unlike the static case, for rotating black hole this  blocked  direction is not the radial one.

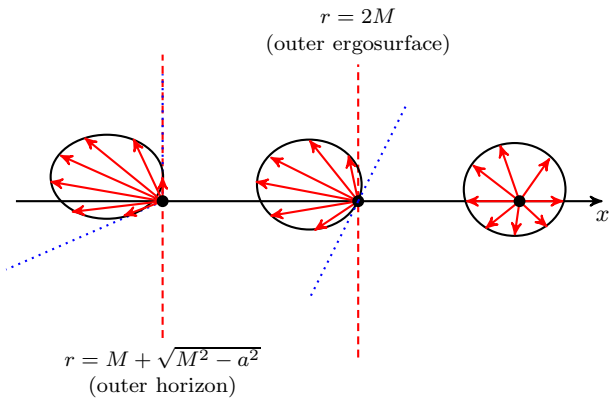
\begin{figure}[t]
\begin{tikzpicture}[
    thick, scale=1.3, 
    >=stealth',
    dot/.style = {
      draw,
      fill = white,
      circle,
      inner sep = 0pt,
      minimum size = 4pt
    }
  ]
  
 \draw[->] (1,0) -- (7,0) coordinate[label = {below:$x$}]  ;  
 
\draw[red,densely dashed] (4.5,-1.6) -- (4.5,1.4) ; 
    
\node  at (4.5,1.9) {\footnotesize $r=2M$};
\node  at (4.5,1.6) {\footnotesize (outer ergosurface)};

\draw[red,densely dashed] (2.5,-1.4) -- (2.5,1.5) ; 
    
\node  at (2.5,-1.6) {\footnotesize $r=M+\sqrt{M^2 - a^2}$};
\node  at (2.5,-1.9) {\footnotesize (outer horizon)};

\draw[dotted,blue] (2.5,0) -- (2.5,1.3) ;
\draw[dotted,blue] (2.5,0) -- (0.9,-0.7) ;

\draw (4.0 ,0.17) ellipse (0.535 and 0.46) ;

\draw[->,red] (4.5,0) -- (4.4,0.47);
\draw[->,red] (4.5,0) -- (4.0,0.625);
\draw[->,red] (4.5,0) -- (3.6,0.45);
\draw[->,red] (4.5,0) -- (3.45,0.2); 

\draw[->,red] (4.5,0) -- (3.6,-0.15); 
 \draw[->,red] (4.5,0) -- (4.05,-0.3);

\filldraw  (4.5,0) circle (0.05);

\draw[dotted,blue] (4.5,0) -- (5,1) ;
\draw[dotted,blue] (4.5,0) -- (4,-1) ;


\draw (6.1,0.12) ellipse  (0.52 and 0.475) ;

\draw[->,red] (6.15,0) -- (6.6,0);
\draw[->,red] (6.15,0) -- (5.6,0);

\draw[->,red] (6.15,0) -- (5.67,0.37);
\draw[->,red] (6.15,0) -- (5.8,-0.28);

\draw[->,red] (6.15,0) -- (5.95,0.575);
\draw[->,red] (6.15,0) -- (6.1,-0.35);

\draw[->,red] (6.15,0) -- (6.46,0.44);
\draw[->,red] (6.15,0) -- (6.435,-0.265);

\filldraw  (6.15,0) circle (0.05);


\draw (1.93,0.25) ellipse (0.575 and 0.43) ; 

\draw[->,red] (2.5,0) -- (2.5,0.25);
 
\draw[->,red] (2.5,0) -- (2.2,0.64);
\draw[->,red] (2.5,0) -- (1.7,0.65);
\draw[->,red] (2.5,0) -- (1.45,0.47);
\draw[->,red] (2.5,0) -- (1.35,0.2);
\draw[->,red] (2.5,0) -- (1.57,-0.11);

\draw[->,red] (2.5,0) -- (2.1,-0.15);

\filldraw  (2.5,0) circle (0.05);
\end{tikzpicture}
    \caption{Depiction of the Finslerian indicatrices for 
the ergosphere and the outer event horizon 
along the $x$ axis in the equatorial plane of a Kerr black hole ($a>0$) in Natario's    coordinates (\ref{Nat}).  When $a<0$,   the figure is reflected with respect to the $x$ axis, corresponding to the transformation $y\to -y$ in (\ref{KerrIndCart}).  In all cases, the indicatrices are horizontal ellipses, with their centers displaced according to   (\ref{KerrIndCart}). 
Notice that the centers of the  indicatrices are also displaced in the azimuthal direction (vertical in the figure) due to the well-known frame-dragging effect.        
     Right: The weak drift regime, in which null geodesics starting from a given point can propagate in all directions. 
      Center: The critical regime corresponding to the ergosurface. The propagation of null geodesics is restricted to directions that are positively aligned with the drift, as shown in equation (\ref{halfspace}).
In contrast with the Schwarzschild ($a=0$) case, the half-space of possible propagation directions (bounded by the blue dotted line) is not tangent to the ergosphere, meaning that
null geodesics starting on the ergosphere can still escape to infinity. Nevertheless, they are subjected to
$d\phi_{BL}\ge 0$, meaning that
null geodesics are corotating with the black-hole in  the usual Boyer-Lindquist   coordinates. This is the only case where we can
have  blocked  directions $(dx^i=0)$.     
       Left: The event horizon. In this case, the propagation is similarly limited to the directions that are positively aligned with the drift and, moreover, constrained to lie within the conic region outlined in   (\ref{cone}) (blue dotted line). Notice the cone is
       such that $dr\le 0$ on the event horizon, meaning that no null geodesics can escape from the interior of the horizon.   This is a concrete realization, for 
a  Kerr black hole, 
  of the wind-Finslerian description 
 of the causal structure  of spacetimes introduced in \cite{miguel}.  
    }\label{fig2}
\end{figure}

Fig.~\ref{fig2} illustrates the results above for geodesics confined to the equatorial plane $\theta = \pi/2$
and starting along the $x$ axis. 
In this case, the indicatrix equation (\ref{KerrInd})    reads
\begin{eqnarray}
\label{KerrIndCart}
&&
\frac{r^2+a^2}{\Sigma}\left(dx + \sqrt{\frac{2M}{r}} \right)^2 \nonumber \\ && +\frac{\Sigma}{r^2+a^2} \left(d y - \frac{2Ma\sqrt{r^2+a^2}}{r\Sigma}  \right)^2 = 1,  
\end{eqnarray}
where we have used explicitly the fact that Natario's coordinates are also of the oblate spheroidal type and, hence, on the equatorial plane the relations $r^2 + a^2 = x^2 + y^2$ and $\tan \phi = y/x$ hold.
The solutions of the equation above, corresponding to the indicatrices for rotating black holes, are ``horizontal'' ellipses with displaced center,  analogously to the
static case of the previous section. Moreover, the closer the starting point of the null geodesics is to the ring singularity at $r=0$, the more elongated the ellipse becomes in the $x$ direction. Furthermore, in contrast with the previous static cases, there is now also a displacement in the vertical direction, which  is clearly related to the   well-known frame-dragging effects.

It is convenient to consider the azimuthal Boyer-Lindquist angle (\ref{phiBL}) to understand the distinctive properties of the ergosurfaces and their interior region.  
For the  blocked geodesic described by (\ref{align}), we have
\begin{equation}
\label{ergoBL}
d\phi_{BL} =\alpha\Omega \left(1+\frac{\rho^2v^2}{\Delta} \right),
\end{equation}
demonstrating that the ergosurface corresponds to the frontier of the ``corotating'' region (ergosphere) where $d\phi_{BL}$ and $\Omega$ have the same sign, {\em i.e.}~$d\phi_{BL}\ge 0 \, (\le 0)$ for $\Omega>0 \, (<0)$. Notice   (\ref{ergoBL}) 
explicitly shows the
well-known
 fact that the azimuthal Boyer-Lindquist angle is well defined only in the exterior region of the
event horizon, in clear contrast with Natario's azimuthal coordinate.

The limit of small angular momentum is particularly interesting from a physical perspective. In this limit, where only linear terms in 
$a$ are considered, the spatial metric   (\ref{Nat}) simplifies to a flat metric. As a result, we recover the Zermelo navigation problem associated with the slowly rotating Kerr metric, as discussed in reference \cite{LiJia}. In this scenario, the Finslerian indicatrices are qualitatively similar to those shown in Fig. \ref{fig2}, with the event horizon and the ergosurface positioned very close to each other.

  Finally, it is noteworthy that the Finslerian indicatrices depicted in Fig.~\ref{fig2}, particularly in the critical and strong drift cases, suggest potential interpretations of superradiance and the Penrose process through the Finslerian framework. In simple terms, these intimately connected phenomena \cite{superradiance} allow for the extraction of energy from Kerr black holes. Superradiance is a scattering phenomenon in which outgoing waves carry more energy than the incoming ones. Similarly, the Penrose process typically involves particle collisions near the horizon, where the black hole absorbs a particle with negative energy, allowing a more energetic particle to escape to infinity The existence of the ergosphere, where no observer can remain static, is crucial for both phenomena.  We have established that the ergosurface $r=2M$, which forms the external boundary of the ergosphere, corresponds to the critical drift region in Natario's coordinates.

 Nevertheless, it is important to note that superradiance can also occur in flat spacetimes, even in the absence of horizons or an ergosurface~\cite{superradiance,nohorizon}. A particularly illustrative example is the Vavilov-Cherenkov effect, in which radiation is emitted by a charge moving faster than the speed of light in a dielectric medium. This phenomenon, which has a long and rich history~\cite{hist1,hist2}, can be explained in terms of spontaneous superradiance involving the formation of a Mach-like cone that characterizes a shock wave~\cite{superradiance}. The shock waves associated with superluminal motion in a dielectric can be understood as a superradiant amplification of electromagnetic waves \cite{hist3}, and similar effects can also be observed with sound waves and gravitational interactions \cite{bekschiff,saaschiff}.  In all these cases, the Mach-like cone has an opening angle
 \begin{equation}
 \label{MC}
 \vartheta = \pi - 2\cos^{-1}\left( \frac{v_{ph}}{v}\right) = 2\cos^{-1}\sqrt{ 1-\left(\frac{v_{ph}}{v}\right)^2} ,
 \end{equation}
 where $v_{ph}$ denotes the phase velocity of the radiation in the external medium and 
 $v = ||\mathbf{V}||$, with $\mathbf{V}$ being the particle velocity.  

The key observation here is that the Vavilov-Cherenkov effect can also be qualitatively understood from the comoving reference frame, where the particle is at rest and the medium is moving with velocity $-\mathbf{V}$.   We can follow    \cite{Unruh} and
describe the Vavilov-Cherenkov radiation in terms of plane waves moving in a medium with phase velocity $v_{ph}$
\begin{equation}
\varphi_\mathbf{k}(t,\mathbf{r})= e^{-i\left(   v_{ph} k t - \mathbf{k}\cdot \mathbf{r} \right)} ,
\end{equation} 
where    $\mathbf{k}$ stands for the wave vector  and $k = ||\mathbf{k}||$. From the comoving reference frame,
such modes are perceived as harmonic excitations $\varphi_\mathbf{k}(t(\tau),\mathbf{r}(\tau)) = e^{-i\omega \tau}$ with frequencies 
\begin{equation}
\label{omega1}
\omega = \gamma\left( v_{ph} k -  \mathbf{V}\cdot \mathbf{k}\right) ,
\end{equation}
 where $\tau$ is the comoving proper time and   
  $\lambda$ is the Lorentz factor. The  Vavilov-Cherenkov effect is associated with the negative frequencies inside the  Vavilov-Cherenkov cone, see \cite{Unruh}  for further details. It turns out that the negative frequencies $\omega$ in (\ref{omega1}) occur exactly inside the cone defined by (\ref{MC}). 
It is easy to see that,
for the cases with $v > v_{ph}$,  
  in the comoving   frame,   the modes emitted inside the cone (\ref{MC}) have negative frequency (energy), while those emitted outside the cone have positive frequency.  Superradiant excitations are known to arise from the energy balance involving positive and negative frequency modes
 \cite{Unruh}. However, from the comoving reference, light rays in the external moving medium will behave locally exactly as predicted by the wind-Finslerian indicatrices with $\mathbf{W}=-\frac{1}{v_{ph}}\mathbf{V}$. In other words, once the critical drift region is crossed and the directions of the null geodesics become constrained by the cone structures described in equations (\ref{halfspace}) and (\ref{cone}), the situation closely resembles that of spontaneous superradiance in flat spacetimes.  From (\ref{cone}), with the standard definition of the angle between two vectors in terms of their scalar product, the opening angle of the Finsler cones is given by
 \begin{equation}
 \vartheta =   2\cos^{-1}\sqrt{1 - \frac{1}{||W||^2}},
 \end{equation}
 which coincides with (\ref{MC}) in the present case.  
  As expected, the cone forms at the critical region where $||W||=1$,
 and becomes more acute as we plunge into the strong drift region 
$||W||>1$. For static black holes, the critical region coincides with the horizon and, as a result, any potential superradiance arising from   negative energy radiation emitted within the cone will remain confined to the interior of the black hole. In contrast, for stationary spacetimes such as the Kerr black hole spacetime, the critical region corresponds to the ergosurface, which lies outside the event horizon. This creates a region, the ergosphere, from which radiation emitted within the cone can escape to infinity  (see Fig. \ref{fig2}), potentially leading to superradiance. This scenario paves the way for a new and interesting interpretation  of superradiance in terms of the  wind-Finslerian  structure of black holes.

\section{Analog models}
\label{sec4}

Analog gravity is a robust framework that explores the emergence of effective curved spacetime geometries in non-gravitational systems~\cite{Barcelo:2005fc,Richartz:2009mi,Jacquet:2020bar,Almeida:2022otk}. These analog spacetimes are particularly suitable for analysis from a  wind-Finslerian    perspective, as they often involve effective metrics of the Painlevé-Gullstrand form (\ref{PG}). For many analog spacetimes, such as those based on hydrodynamic flows and Bose-Einstein condensates, the ``river model" description is not merely an alternative interpretation, but rather the genuine underlying physics of the system. 

An especially illustrative class of analog models where the Finslerian formalism finds direct and novel application is that of  {vortex flows} in  fluids, often employed as analogs of rotating black holes.
This is the case, for instance, of the configuration known as the ``draining bathtub'' model, described in the standard Euclidean two-dimensional space  
by the  background velocity field 
\begin{equation}
\label{vortex}
 {W} = -\frac{A}{r}\hat{r} + \frac{B}{r}\hat{\varphi} ,
\end{equation} 
  where the radial component $A>0$ controls the inward drift and $B$ encodes the vortex circulation. The associated acoustic metric, derived via linear perturbations of the velocity potential, has a Painlevé--Gullstrand  form (\ref{PG}), see \cite{DBT} for details.  In this case,
\begin{equation}
||W||^2 = \frac{A^2 + B^2}{r^2},
\end{equation}
and the Finslerian indicatrix for null geodesics starting along the $x$ axis reads
\begin{equation}
\label{DBT}
\left(dx +  {\frac{A}{r}} \right)^2  + \left(dy -  {\frac{B}{r}} \right)^2  = 1.
\end{equation}
The overall behavior of this indicatrix is qualitatively similar to those of the previous section. The
 {strong drift regime} emerges where the inward radial flow exceeds the local wave propagation speed $c_w=1$, resulting in a trapping horizon. This occurs at $r_h=A$, where the underlying geometry is governed by a 
  {Lorentz-Finsler metric}. On the other hand, an analog ergosurface is found at the critical region  
  $r_e = \sqrt{A^2+B^2}$, where $||W||=1$.
Thus, the presence of the {azimuthal component of the drift} leads to the formation of an  {ergoregion}
$
r_h < r < r_e
$, analogous to that of a Kerr black hole, where the co-rotating frame speed exceeds the local wave speed in the azimuthal direction. 

This analog ergoregion possesses the necessary geometric features for superradiant scattering, and, indeed, an active area of research with experimental results~\cite{nature} has developed around this topic. 
Whenever an ergoregion is present together with a dissipation mechanism, wave modes with azimuthal number $m$ and frequency $\omega$ satisfying $\omega < m\Omega$, where $\Omega$ is the local angular velocity of the flow, are amplified upon reflection due to the presence of negative-energy modes in the comoving frame \cite{Barcelo:2005fc,DBT,Richartz:2009mi}. The  {Finsler indicatrix} (\ref{DBT}), being displaced and skewed by the azimuthal drift as in Fig.~\ref{fig2}, encodes this kinematic conditions geometrically, reflecting the frame-dragging and the energy  balance that underpins superradiance. In this way,   Finsler geometry offers a covariant and geometrically transparent description of the interplay between ergoregion formation and 
superradiance 
 phenomena in vortex-based analog black hole systems.  Furthermore, this framework may be instrumental in analyzing stability properties and backreaction effects in experiments involving dispersive corrections.
 
We finish by commenting that flows 
of the type (\ref{vortex}),
 with $A<0$ and  $B=0$, correspond to the standard case of the
 circular hydraulic jump,  a canonical example of an analog white hole horizon. Hydraulic jumps form when the propagation of surface waves on a thin liquid layer undergoes a transition from a supercritical to a subcritical regime, marking a causal boundary for counterstream wave propagation~\cite{Froude}.
 In the shallow-water limit, the dispersion relation for gravity surface waves reduces to a relativistic form \cite{Dispersion,Schutzhold:2002rf}
\begin{equation}
\label{disp}
(\omega - \mathbf{W} \cdot \mathbf{k})^2 \approx  ||\mathbf{k}||^2 = k^2,
\end{equation}
where $\mathbf{W}$ is the  local background flow velocity, and $\omega$ and $\mathbf{k}$ 
denote the  wave frequency and the wave vector, respectively. It turns out that the dispersion relation
(\ref{disp}) admits an unexpected but clear interpretation in Finslerian terms. Notice that a wave traveling in two dimensions, 
with frequency $\omega$ and
wave vector $\mathbf{k}$, 
 moves along the vector $\mathbf{V} = dx\hat\imath + dy \hat \jmath = \frac{\omega }{k^2}\mathbf {k}$. In terms of the vector
 $\mathbf{V}$, for a radial flow $\mathbf{W} = W\hat r$, with $W= \frac{|A|}{r} > 0$, we find
\begin{equation}
\left(dx -W \right)^2 + dy^2 = 1
\end{equation} 
for waves starting on the $x$ axis. Thus, the dispersion relation (\ref{disp}) defines an indicatrix, exhibiting the same qualitative behavior discussed in the previous sections. The critical region $r_h=|A|$ acts as a white hole horizon and determines
the location of the hydraulic jump of the flow~\cite{Froude}. Outside the hydraulic jump, 
we are in the weak drift regime, where 
surface waves can propagate in all directions. 
 At the radius of the hydraulic jump $r = R_j$, the drift matches the wave speed so that only radially outward propagation is permitted. 
Finally, in the strong drift regime, corresponding to the interior of the hydraulic jump, 
only outgoing geodesics exist, mimicking the causal disconnection characteristic of white hole interiors. As one can see, the {\color{blue}wind-Finslerian} picture offers a rigorous and geometrically transparent framework for describing horizon formation, causal disconnection, and wave dynamics 
also for analog models involving hydraulic jumps.

\section{Final remarks}

We have shown, through explicit and relevant physical examples, that recent mathematical findings (see \cite{miguel} for the main
results and a 
comprehensive 
 review) regarding the Zermelo navigation problem and its underlying   wind-Finslerian structure  can provide a complete description of the causal structure of black holes, answering affirmatively the question posed in Ref.~\cite{LiJia} about the extension of the Finsler structure to the interior of black holes. 
  Such results are a direct consequence of the  wind-Finslerian structures introduced and developed in \cite{miguel}. 
 Specifically, the Kropina metric, associated with the critical Zermelo navigation problem, can fully characterize the causal properties of ergosurfaces in stationary black holes
and event horizons in the spherically symmetric case. On the other hand, the Lorentz-Finsler metric, which applies to the case of a strong Zermelo problem, describes the causal structure of the interior (non-stationary) region of black holes. This complete   wind-Finslerian  description offers an enhanced visual understanding of the river model of black holes, presenting the behavior of null geodesics on the ergosurfaces, horizons, and their interior regions in novel geometric terms. There are many other recent mathematical results that may have physical applications within this framework; these topics are currently under investigation.

We have two final remarks. The first one is related to Fermat's principle, which is normally invoked to motivate physically, or even
justify,  
the fact that the null geodesics should correspond to minimal time paths such that $dt=F(\mathbf{x},d\mathbf{x})$. Since
$g_{00} = 1 - ||W||^2$ for the Painlevé-Gullstrand metric (\ref{PG}), we conclude that $dt$ is timelike only in the
weak drift region, or outside the ergosurface/horizon 
in the stationary/static case. This can raise   questions on the interpretation, or even the validity,
 of the Finslerian geodetic 
condition for  light propagation. 
We stress that our approach is consistent and valid  regardless of the  nature of the coordinate $t$ of the Painlevé-Gullstrand metric.
The coordinate $t$ must be interpreted as a parameter for the null geodesics, which happens to be associated with a timelike direction in the weak drift case. Considered as a parameter, we  have $F\left( \mathbf{x}, \frac{d\mathbf{x}}{dt}\right)=1$, which is a differential equation for the geodesic $x^i(t)$. Our approach is based on the mathematical fact that this differential equation is equivalent to the geodesic (Euler-Lagrange) equations of the action $S=\int F(x^i,\dot x^i) dt$. Since $F(x^i,\dot x^i)$ is positively homogeneous of
degree one, its Hamiltonian is identically zero. To avoid unnecessary issues involving constraints, one can consider the equivalent
action associated with the Lagrangian $L(x^i,\dot x^i) = F^2(x^i,\dot x^i)$ (see, for instance,  \cite{{gibbons2009stationary}}). Since $L(x^i,\dot x^i)$ does not depend explicitly on $t$, its associated Hamiltonian function $H$ is conserved. However, it turns out that $H=L$ in our case, and hence
$F(x^i,\dot x^i)$ is constant on the solutions. In summary, irrespective of the nature of the parameter $t$, the null geodesics are always the solution of the Euler-Lagrange equations of $S=\int F^2(x^i,\dot x^i) dt $ with $H=1$.  Needless to say, such heuristic arguments are fully compatible with the rigorous definitions and results of wind-Finslerian structures of Ref. \cite{miguel}. 

 Our second and final remark concerns the analog gravity program. Besides the issues discussed in Section \ref{sec4},  the study of Finsler geometries can be further advanced by investigating their connections to analog spacetimes, such as those realized in hydrodynamic flows~\cite{gibbons2009stationary,App9,Barcelo:2001cp,Visser:2007du}, Bose-Einstein condensates~\cite{Barcelo:2001cp,Visser:2007du,Weinfurtner:2006wt}, and birefringent crystals~\cite{App4,Skakala:2008kf}. As we have argued, the interplay between Spacetime Physics and Finsler geometries becomes particularly evident when the metric is recast in the Painlevé-Gullstrand form. The Painlevé-Gullstrand form not only emphasizes the causal structure of the underlying spacetime, but also naturally aligns with the   wind-Finslerian  framework. In analog spacetimes, this alignment is not incidental since the Painlevé-Gullstrand form is particularly suitable for systems with a preferred frame, such as fluids or optical media. By establishing a direct connection between Finsler geometries and the causal structure of black hole interiors through explicit constructions, our work lays the foundation for deeper investigations into analog horizons and ergosurfaces, offering fresh perspectives on  wind-Finslerian  concepts within the framework of analog gravity.

\section*{Acknowledgments}
This study was financed in part by the São Paulo Research Foundation (FAPESP, Brazil) - Process Numbers 2022/15371-3 (HR), 2022/08335-0 (MR),  2021/09293-7 (AS). The authors also acknowledge partial support from Conselho Nacional de Desenvolvimento Científico e Tecnológico (CNPq, Brazil), Grants 315991/2023-2 (MR) and 306785/2022-6 (AS). 
AS 
also wishes to thank Vitor Cardoso and José S. Lemos for
the warm hospitality at the Center for Astrophysics and
Gravitation of the University of Lisbon, where part of this  work
was done, and S. Carneiro for  helpful   discussions.

\end{document}